\documentclass[conference]{IEEEtran}
\IEEEoverridecommandlockouts
\usepackage[utf8]{inputenc}
\usepackage{graphicx}
\usepackage{amsmath}
\usepackage{amsfonts}
\usepackage{amssymb}
\usepackage{booktabs}
\usepackage{multirow}
\usepackage{siunitx}
\usepackage{cite}
\usepackage{url}
\usepackage{xcolor}

\newcommand{\bh}[1]{\textbf{\boldmath #1}}

\title{2.4-GHz Integrated CMOS Low-Noise Amplifier (English Version*)
\thanks{*This document is the author's translation of a peer-reviewed paper published initially in Spanish. \textbf{How to cite}: J. L. Gonz\'alez, J. C. Cruz, R. L. Moreno, and D. V\'azquez, ``2.4-GHz Integrated CMOS Low-Noise Amplifier," in V International Symposium on Electronics, XVI Convention Informatica 2016, La Habana, Cuba, 14-18 Mar, 2016.}
}

\author{\IEEEauthorblockN{Jorge L. Gonz\'alez-Rios, Juan C. Cruz Hurtado, Robson L. Moreno, Diego V\'azquez}
}

\begin{document}
\maketitle

\begin{abstract}
This paper presents the analysis, design, fabrication, and measurement of an integrated low-noise amplifier (LNA) implemented using a 130 nm CMOS technology, operating in the 2.4 GHz band. The LNA is a crucial component in the performance of receivers, particularly in integrated receivers. The proposed LNA was designed to meet the specifications of the IEEE 802.15.4 standard. Post-layout simulation results, including pads with electrostatic discharge (ESD) protection, are as follows: gain of 10.7 dB, noise figure of 2.7 dB, third-order input intercept point (IIP3) of 0.9 dBm, input and output impedance matching better than -20 dB with respect to 50~$\Omega$ terminations, with a power consumption of 505 $\mu$W powered from a 1.2 V supply. The obtained results fall within the range of those recently reported for the same topology and operating frequency. The measured scattering parameters (S-parameters) are consistent with the simulation results. This work contributes to the development of a new research line in Cuba on the design of radio-frequency (RF) integrated circuits.
\end{abstract}


\section{Introduction}
\label{sec:intro}
Short-range wireless communication devices are increasingly offering a higher number of applications, which pose design challenges such as miniaturization, low-voltage operation, and reduced energy consumption. These demands can be satisfied using current CMOS technologies, which allow the integration of all system blocks, including RF circuits, into a single chip \cite{ref01,ref02}.

Data transfer in these applications requires receivers with a wide dynamic range due to the variability of RF signal levels and the presence of multiple interferers \cite{ref03}. Therefore, multiple design trade-offs arise when implementing the different RF front-end blocks \cite{ref04}, especially in the low-noise amplifier (LNA), which is the first active block of the receiver.

The LNA determines the minimum detectable signal of the receiver, through a sufficiently high gain and a low noise figure \cite{ref04,ref05}. On the other hand, too high a gain can saturate subsequent blocks (such as the mixer) in the presence of strong input signals. Simultaneously, the LNA must also provide good input impedance matching, sufficiently high linearity, and good reverse isolation.

\begin{figure}[!t]
 \centering
 \includegraphics[width=\linewidth]{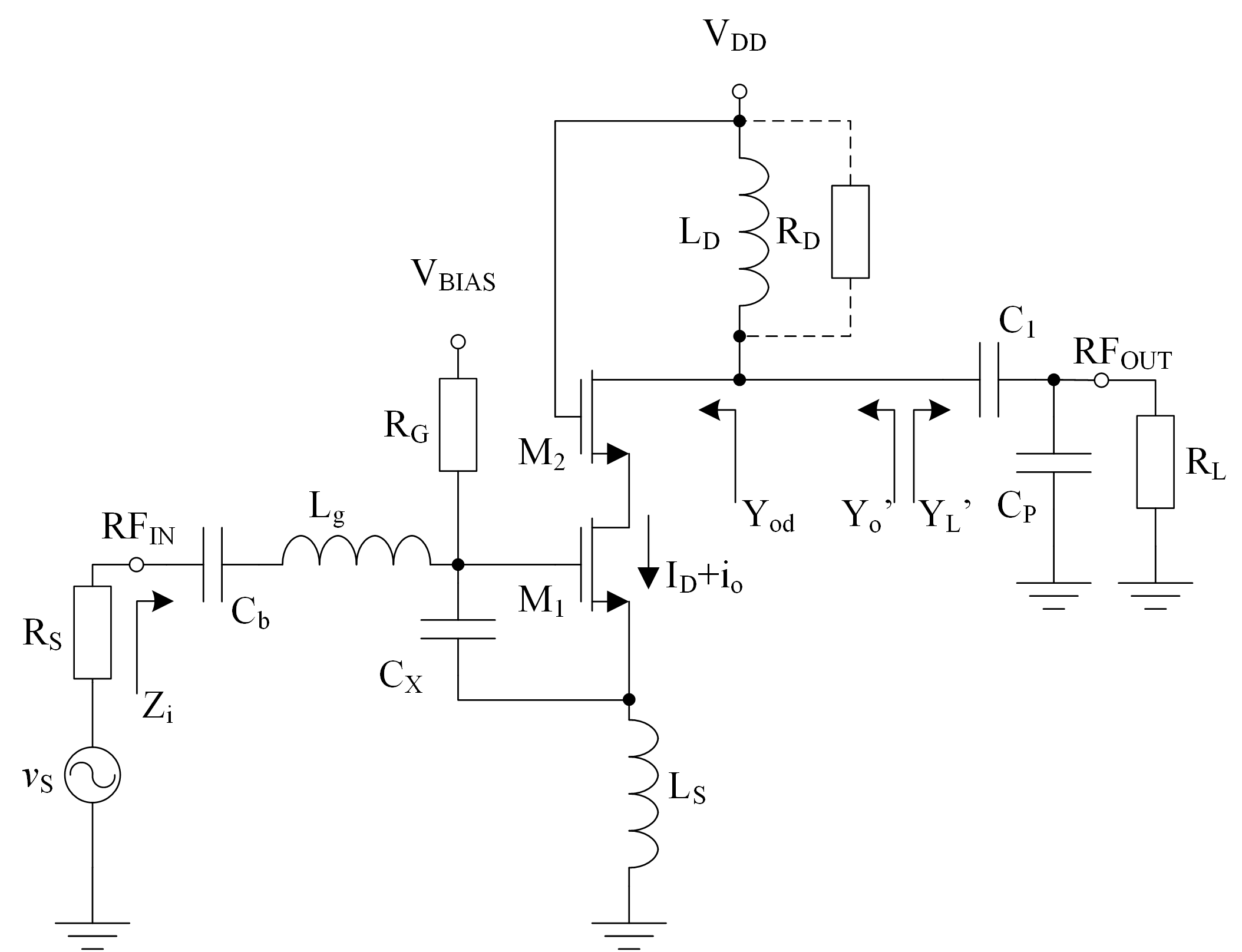}
 \caption{Common-source LNA topology with inductive degeneration.}
 \label{fig1}
\end{figure}

This work presents an approach to the analysis, design, fabrication, and experimental characterization of a CMOS LNA with common-source topology and inductive degeneration (Figure~\ref{fig1}). This topology is widely used in integrated CMOS receivers for short-range communications \cite{ref05,ref06,ref07}. The design was implemented to meet the specifications of the IEEE 802.15.4 standard \cite{ref07} at 2.4 GHz, and fabricated in a 130 nm, 1.2 V CMOS technology. This work contributes to the development of a new research line in Cuba on the design of RF integrated circuits \cite{gonzalez2016ieeelat,gonzalez2016apec}.

The rest of the paper is organized as follows. 
Section \ref{sec:description} describes the chosen topology and the general design methodology. Section \ref{sec:analysis} presents a simplified circuit analysis, providing elements for selecting the passive components. Section \ref{sec:results} presents the simulation results used for design, the measured scattering parameters, and considerations for physical implementation, as well as experimental results. Finally, Section \ref{sec:conclusions} provides conclusions and outlines future work.

\section{Common-Source CMOS LNA Topology with Inductive Degeneration}
\label{sec:description}
The basic schematic of a common-source LNA (CS-LNA) with inductive degeneration is shown in Figure~\ref{fig1}. The inductive degeneration (via $L_s$) produces the resistive component required to match the input impedance to the preceding stage without introducing additional noise \cite{ref08}. The capacitor $C_X$ helps minimize the noise figure for specific gain and power-consumption targets \cite{ref09}. The gate inductor $L_g$ is included to tune the input impedance. Transistor $M_2$ is used as a cascode stage to reduce the Miller effect on $M_1$ and to improve reverse isolation \cite{ref05}. The drain inductor $L_D$ forms a parallel resonant network with the output capacitances of the cascode stage and the impedance seen toward the load. The capacitive divider $(C_1, C_P)$ is included to couple the output impedance to 50~$\Omega$ for stand-alone LNA characterization. Finally, $C_b$ blocks the DC component from the RF source.

Transistor sizing can be used to minimize the noise figure (NF), as demonstrated in prior work on this topology \cite{ref09,ref10,ref11}. High IIP3 with low power consumption can be achieved by exploiting the linearity ``sweet spot'' in MOS transistors biased in moderate inversion \cite{ref12}. This IIP3 peak occurs at approximately the same current density \cite{ref13,ref14}, so linearity can also be maximized through proper transistor sizing.

Guided by the above considerations and the need to minimize power consumption, the design-space exploration consisted in sweeping the bias current ($I_D$) and the transistor width. For each pair $(I_D, W)$, the passive-element dimensions were synthesized to meet the gain and impedance-matching requirements (LNA synthesis) while respecting the technology limits so that all candidate circuits were physically realizable. 

Once the passive elements were fixed, the NF and IIP3 of each synthesized LNA were obtained. The set of results for the different LNAs constitutes a design space from which the final implementation can be selected, balancing the criteria chosen by the designer (in addition to power, noise, and linearity, other aspects such as occupied area and robustness to process variations may be considered), balancing the criteria chosen by the designer (in addition to power, noise, and linearity, other aspects such as occupied area and robustness to process variations may be considered).

\section{Circuit Analysis}
\label{sec:analysis}
The available power gain of the LNA can be expressed as
\begin{equation}
G \triangleq \frac{P_o}{P_{\text{avs}}}
= \frac{I_o^2/(4\,G'_o)}{V_s^2/(4\,R_S)}
= \frac{G_m^2\,R_S}{G'_o},
\label{eq:G_def_cs}
\end{equation}
where $G_m \triangleq \lvert I_o/V_s\rvert$ is the effective transconductance of the input stage under input matching, $R_S$ is the source resistance, and $G'_o \equiv \Re\{Y'_o\}$ is the real part of the output-stage admittance seen by the load.

The output conductance is the parallel combination of the cascode output conductance and the drain-inductor loss, namely
\begin{equation}
G'_o = \Re\{Y_{od}\} + \frac{1}{\omega_0 L_D Q_D},
\label{eq:Go_cs}
\end{equation}
where $\omega_0$ is the operating angular frequency and $Q_D$ is the quality factor of $L_D$.

Using a simplified small-signal model for the input stage (considering in $M_1$ only $C_{gs}$ and the controlled source $i_o=g_m v_{gs}$, and treating $L_S$, $L_g$, and $C_X$ as ideal), the effective transconductance under input matching can be approximated as
\begin{equation}
G_m \simeq \frac{1}{2\,\omega_0\,L_S}.
\label{eq:Gm_cs}
\end{equation}

Equations~\eqref{eq:G_def_cs}–\eqref{eq:Gm_cs} show that the LNA gain exhibits a trade-off between the characteristics of the drain inductor $L_D$ and the source-degeneration inductor $L_S$: the former contributes through the product $L_D Q_D$ (via $G'_o$), and the latter dominates through its inductance. In selecting $L_D$, the frequency response of the output impedance must also be considered. To widen the frequency span over which the output impedance remains adequate, the quality factor of the output resonant network should be reduced; thus, a lower $Q_D$ is desired. To maintain the same contribution of the output network to the gain when decreasing $Q_D$, $L_D$ must be increased proportionally. However, the maximum achievable inductance is limited by both the technological constraints of on-chip inductors and the minimum capacitance in the output resonant network.

\begin{figure}[!t]
 \centering
 \includegraphics[width=.9\linewidth]{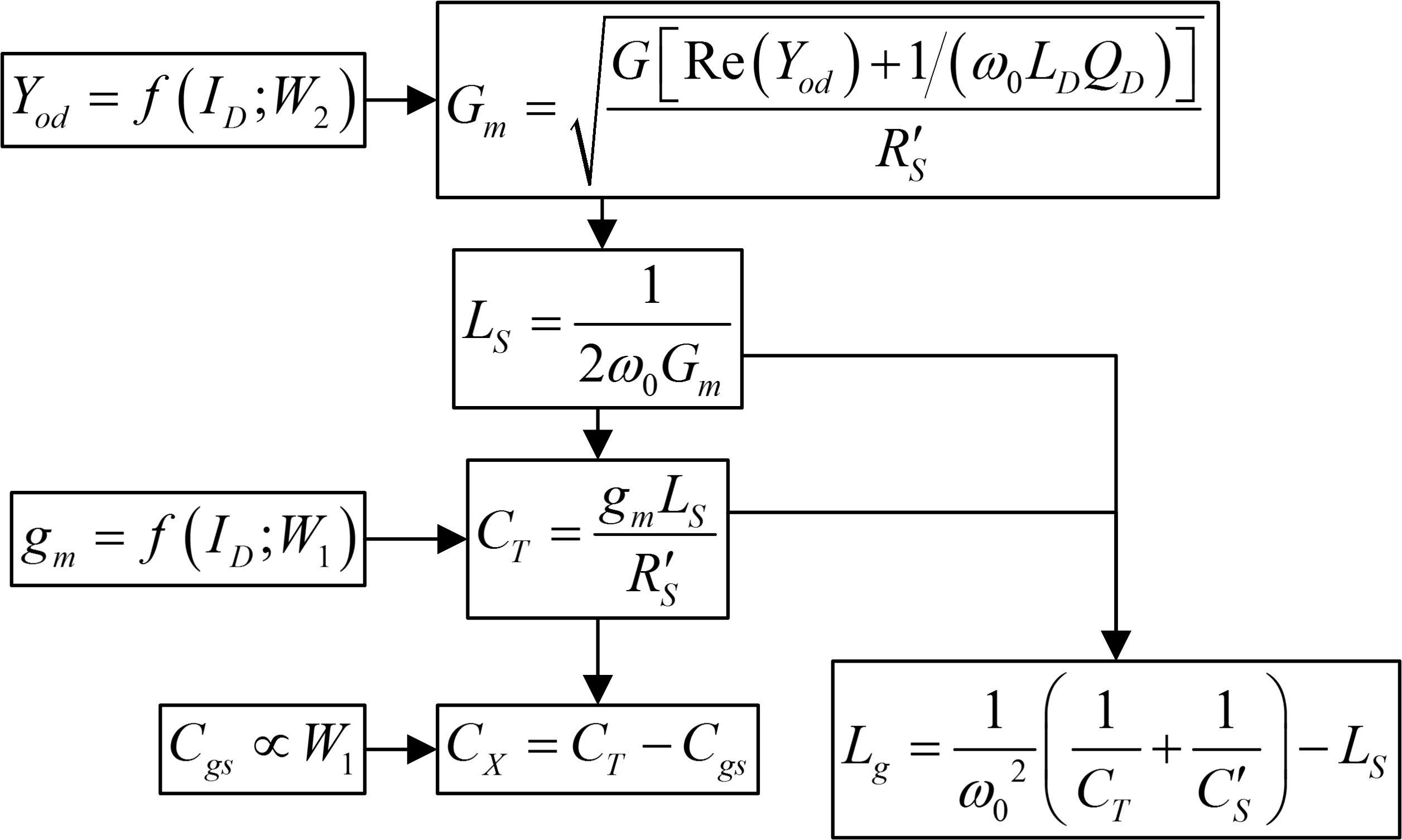}
 \caption{Dependencies of passive elements in the transconductance stage on LNA gain, transistor dimensions, and biasing.}
 \label{fig2}
\end{figure}

Due to these trade-offs in selecting the drain inductor, it is convenient to fix the characteristics of $L_D$ first and then determine the remaining passive elements. The simplified small-signal equations are not accurate enough to directly compute final design values, but they provide a logical sequence for calculating the input-stage passive elements. This sequence is summarized in a flow chart in Figure~\ref{fig2}, which also includes the dependencies on device sizing and biasing. The total equivalent capacitance between the gate and source of $M_1$ is given by $ C_T = C_X + C _ {gs} $.

\section{Results and Discussion}
\label{sec:results}

\subsection{Design Space Exploration}
\label{sec:exploration}
Simulations were performed using the device models provided by a 130 nm CMOS Process Design Kit (PDK) following the RF specifications in Table~\ref{tab1} for ZigBee/IEEE 802.15.4 receivers \cite{ref16,ref17}. Input ($S_{11}$) and output ($S_{22}$) reflection coefficients are referenced to 50~$\Omega$.

\begin{table}[!t]
 \centering
 \caption{LNA Specifications}
 \begin{tabular}{ccccc}
\hline
\textbf{\begin{tabular}[c]{@{}c@{}}Frecuency\\ (GHz)\end{tabular}} & \textbf{\begin{tabular}[c]{@{}c@{}}Gain\\    (dB)\end{tabular}} & \textbf{\begin{tabular}[c]{@{}c@{}}NF\\    (dB)\end{tabular}} & \textbf{\begin{tabular}[c]{@{}c@{}}IIP3\\    (dBm)\end{tabular}} & \textbf{\begin{tabular}[c]{@{}c@{}}S11,   S22\\    (dB)\end{tabular}} \\
\hline
2.4 -- 2.5 & 10.5±0.5 & \textless{}3 & \textgreater{}-4 & \textless{}-10\\
\hline
\end{tabular}
 \label{tab1}
\end{table}

As described in Section~\ref{sec:description}, the exploration consisted of sweeping the bias current ($I_D$) and the width of transistor $M_1$ ($W_1$). The width of $M_2$ was set to $W_2 = W_1/2$ to reduce its contribution to the load capacitance and increase the selection margin of the output matching network \cite{ref06}. The channel length of all transistors was fixed to the minimum allowed by the technology ($L_1 = L_2 = L = 0.12~\mu$m). The supply voltage was $V_{DD} = 1.2$~V.

\begin{figure}[!t]
 \centering
 \includegraphics[width=\linewidth]{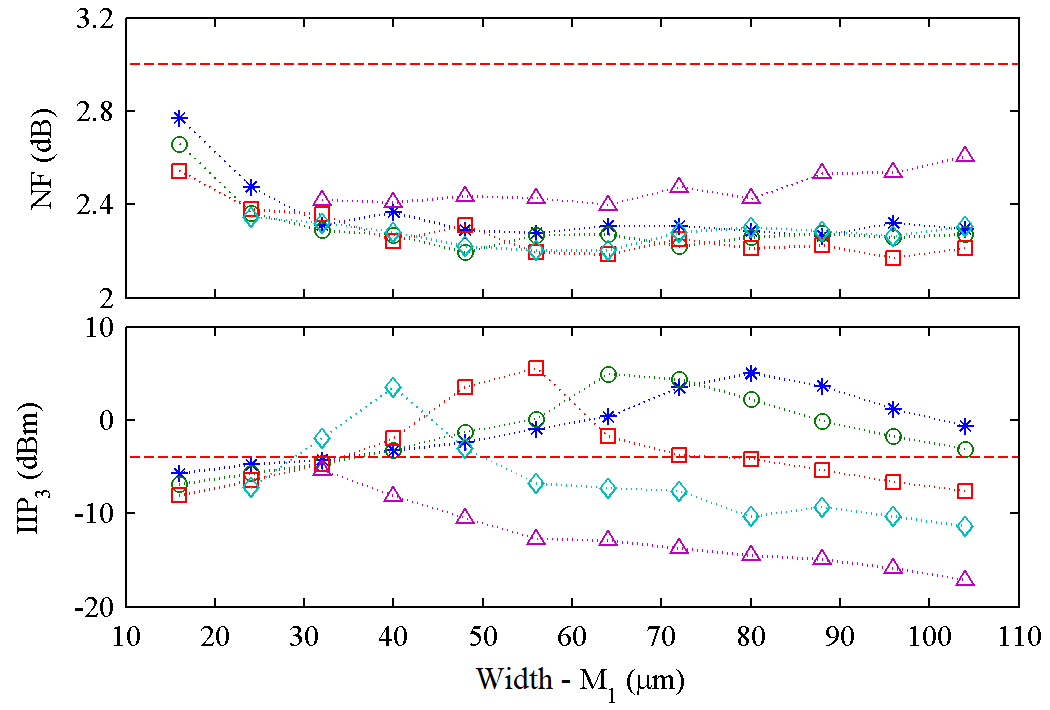}
 \caption{Simulation results at 2.45 GHz for NF (top) and IIP3 (bottom) versus transistor $W_1$ and bias current.}
 \label{fig3}
\end{figure}

Figure~\ref{fig3} shows simulation results at 2.45~GHz for the noise figure (NF) and the input-referred third-order intercept point (IIP3) as a function of $W_1$ and $I_D$. All results shown correspond to circuits with $S_{11}, S_{22} < -15$~dB and gains in the interval $[10.3, 10.9]$~dB. The minimum values of $W_1$ for $I_D=0.4$~mA ($24~\mu$m) and $I_D=0.3$~mA ($32~\mu$m) were limited by the technology constraints on the passive elements required to meet the gain and matching specifications. All synthesized LNAs meet the NF specification ($\text{NF}<3$~dB). However, the required linearity ($\text{IIP3} > -4$~dBm) is not met for $I_D=0.3$~mA, so this bias current must be discarded. Based on these results, we selected the LNA with $I_D=0.4$~mA (the lowest current for which all requirements are met) and $W_1=40~\mu$m, achieving the highest IIP3 at that power level.

\subsection{Fabrication and Experimental Results}
\label{sec:experiment}
\begin{figure}[!t]
 \centering
 \includegraphics[width=.9\linewidth]{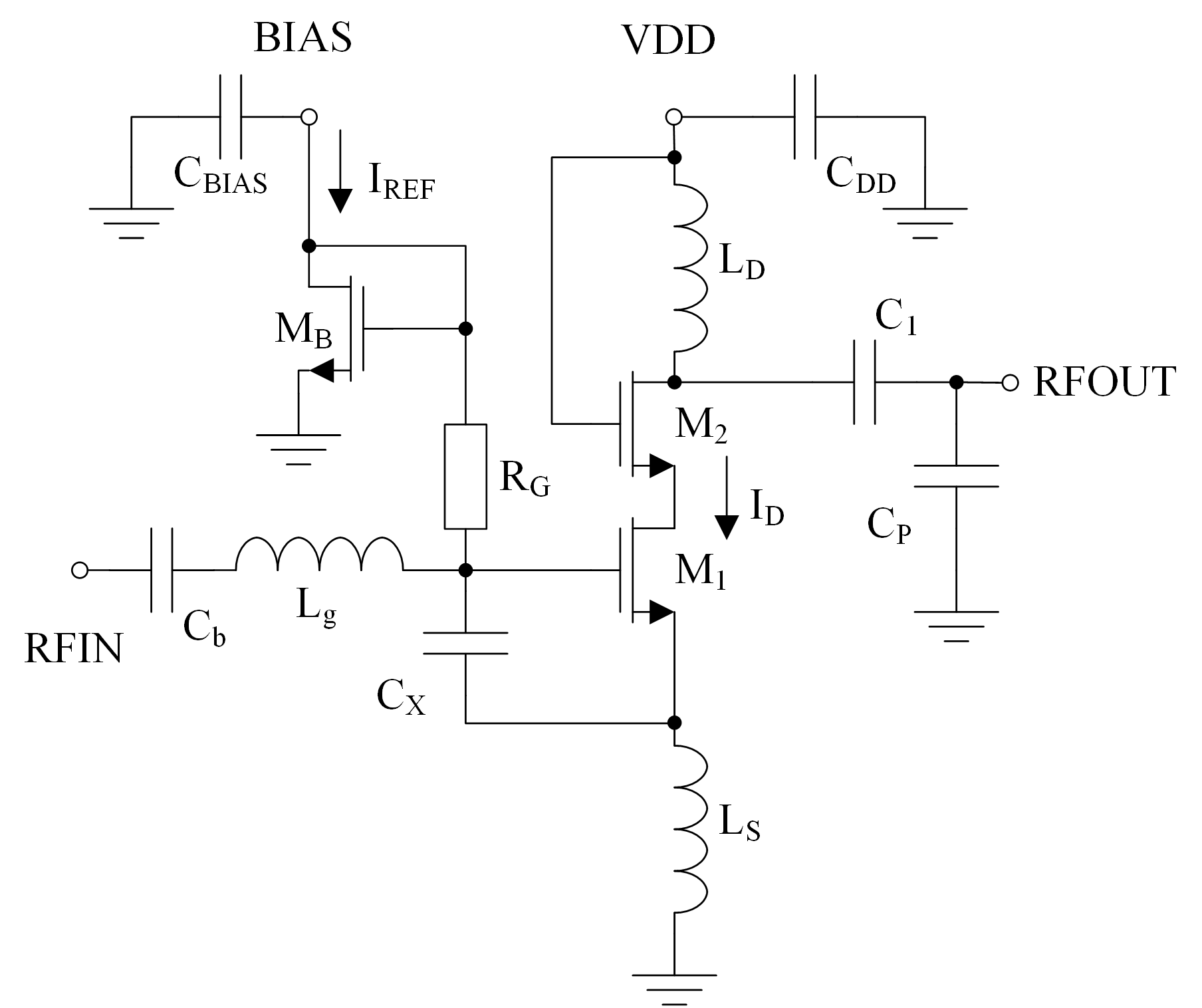}
 \caption{Schematic of the fabricated LNA.}
 \label{fig4}
\end{figure}

Figure~\ref{fig4} illustrates the schematic of the fabricated LNA. The gate of $M_1$ is biased using a current mirror ($M_B$) excited externally with a reference current $I_{\text{REF}}$. All pads are protected against electrostatic discharge (ESD) by diodes \cite{ref06} (not shown in the figure). Decoupling capacitors were included between the bias/supply terminals (BIAS and $V_{DD}$) and ground (GND) to reduce RF coupling into DC lines. Table~\ref{tab2} lists the main device dimensions and component values used (with $V_{DD}=1.2$~V).

\begin{table}[!t]
 \centering
 \caption{Implemented LNA Dimensions}
\begin{tabular}{cccccc}
\hline
\bh{$I_{REF}$} & \bh{$W_1/L_1$} & \bh{$W_2/L_2$} & \bh{$W_B/L_B$} & \bh{$R_G$} & \bh{$C_b$} \\
\textbf{(µA)} & \textbf{(µm/µm)} & \textbf{(µm/µm)} & \textbf{(µm/µm)} & \textbf{(k$\Omega$)} & \textbf{(pF)} \\
\hline
31.5 & 40/0.12 & 20/0.12 & 4/0.12 & 10 & 12.6 \\
\hline
\hline
\bh{$L_S$} & \bh{$C_X$} & \bh{$L_g$} & \bh{$L_D$} & \bh{$C_1$} & \bh{$C_P$} \\
\textbf{(nH)} & \textbf{(fF)} & \textbf{(nH)} & \textbf{(nH)} & \textbf{(fF)} & \textbf{(pF)} \\
\hline
1.8 & 246 & 13.5 & 9.5 & 441 & 1.24\\
\hline
\end{tabular}
 \label{tab2}
\end{table}

\begin{figure}[!t]
 \centering
 \includegraphics[width=\linewidth]{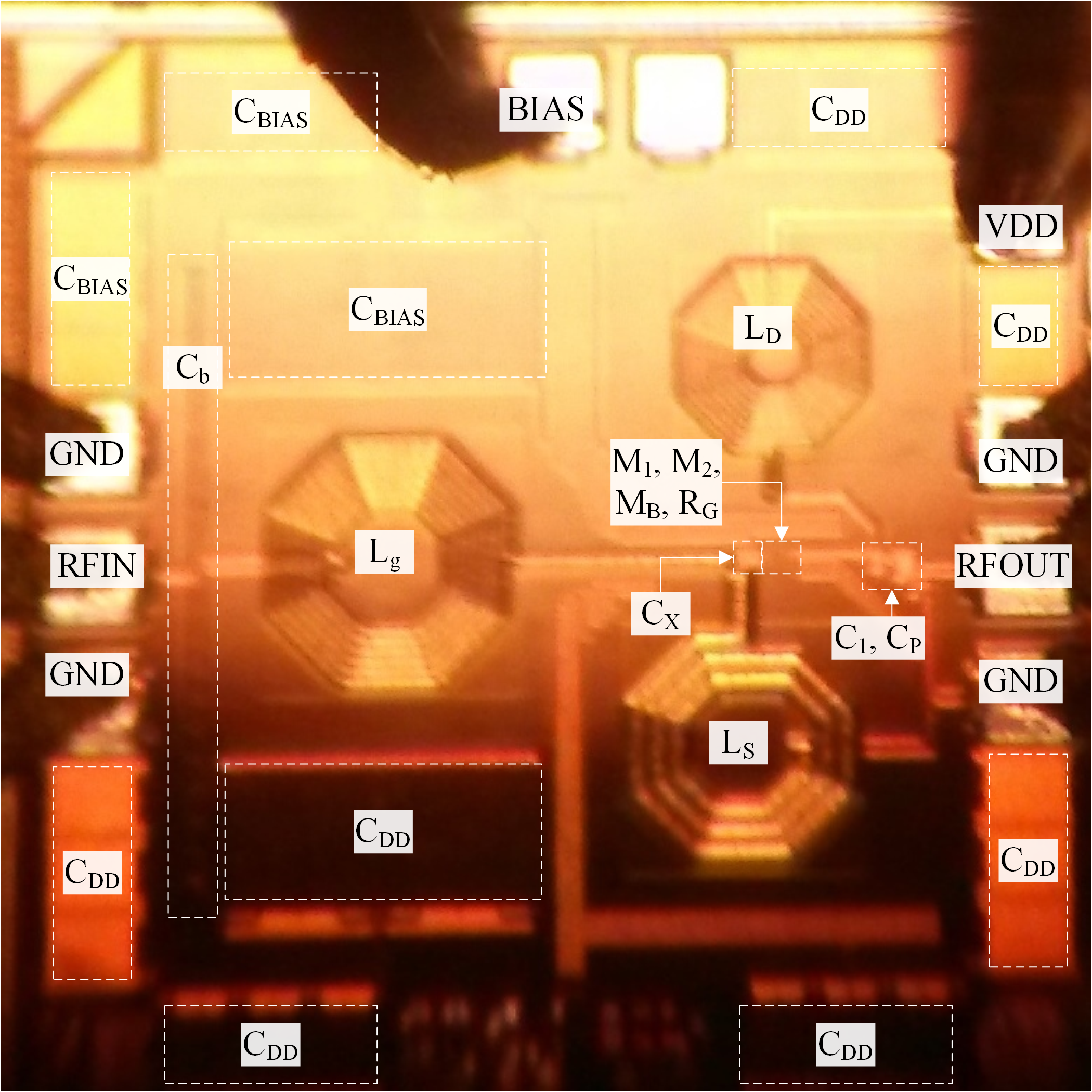}
 \caption{Microphotograph of the fabricated LNA.}
 \label{fig5}
\end{figure}

A microphotograph of the fabricated LNA is shown in Figure~\ref{fig5}. In the implemented layout, the inductors were placed as far apart as possible from each other and from the RF and bias signal traces within the available chip area to minimize electromagnetic coupling, which can affect circuit performance. The free area between pads and between the gate inductor and the bias lines was used to implement large decoupling capacitors (e.g., $C_{DD}=747$~pF and $C_{BIAS}=427$~pF). In the implemented layout, the inductors were placed as far apart as possible from each other and from the RF and bias signal traces within the available chip area to reduce electromagnetic coupling, which affects the circuit performance \cite{ref18,ref19}. The free area between pads and between the gate inductor and the bias lines was used to implement large decoupling capacitors (e.g., $C_{DD}=747$~pF and $C_{BIAS}=427$~pF). The chip area is $1230\times1240~\mu\text{m}^2$, including pads.

Post-layout simulation results (i.e., including parasitic extraction of resistive and capacitive effects) are summarized in Table~\ref{tab3} for the center frequency of interest (2.45~GHz). The total supply current is $I_{DD}=I_{\text{REF}}+I_D$, and $P_{DC}=I_{DD}V_{DD}$. The table demonstrates compliance with all specifications, with power dissipation and RF metrics aligning with the state of the art \cite{ref20}.

\begin{table}[!t]
 \centering
 \caption{Post-Layout Simulation Results}
 \begin{tabular}{c c c c c c c c}
 \hline
 \bh{$I_{\mathrm{DD}}$} & \bh{$P_{\mathrm{DC}}$} & \bh{$G$} & \bh{$NF$} & \bh{$IIP_3$} & \bh{$S_{11}$} & \bh{$S_{22}$} & \bh{$S_{12}$} \\
 \textbf{(mA)} & \textbf{($\mu$W)} & \textbf{(dB)} & \textbf{(dB)} & \textbf{(dBm)} & \textbf{(dB)} & \textbf{(dB)} & \textbf{(dB)} \\
 \hline
 0.42 & 505 & 10.7 & 2.7 & 0.9 & -24 & -30 & -41 \\
 \hline
 \end{tabular}
 \label{tab3}
\end{table}

\begin{figure}[!t]
 \centering
 \includegraphics[width=\linewidth]{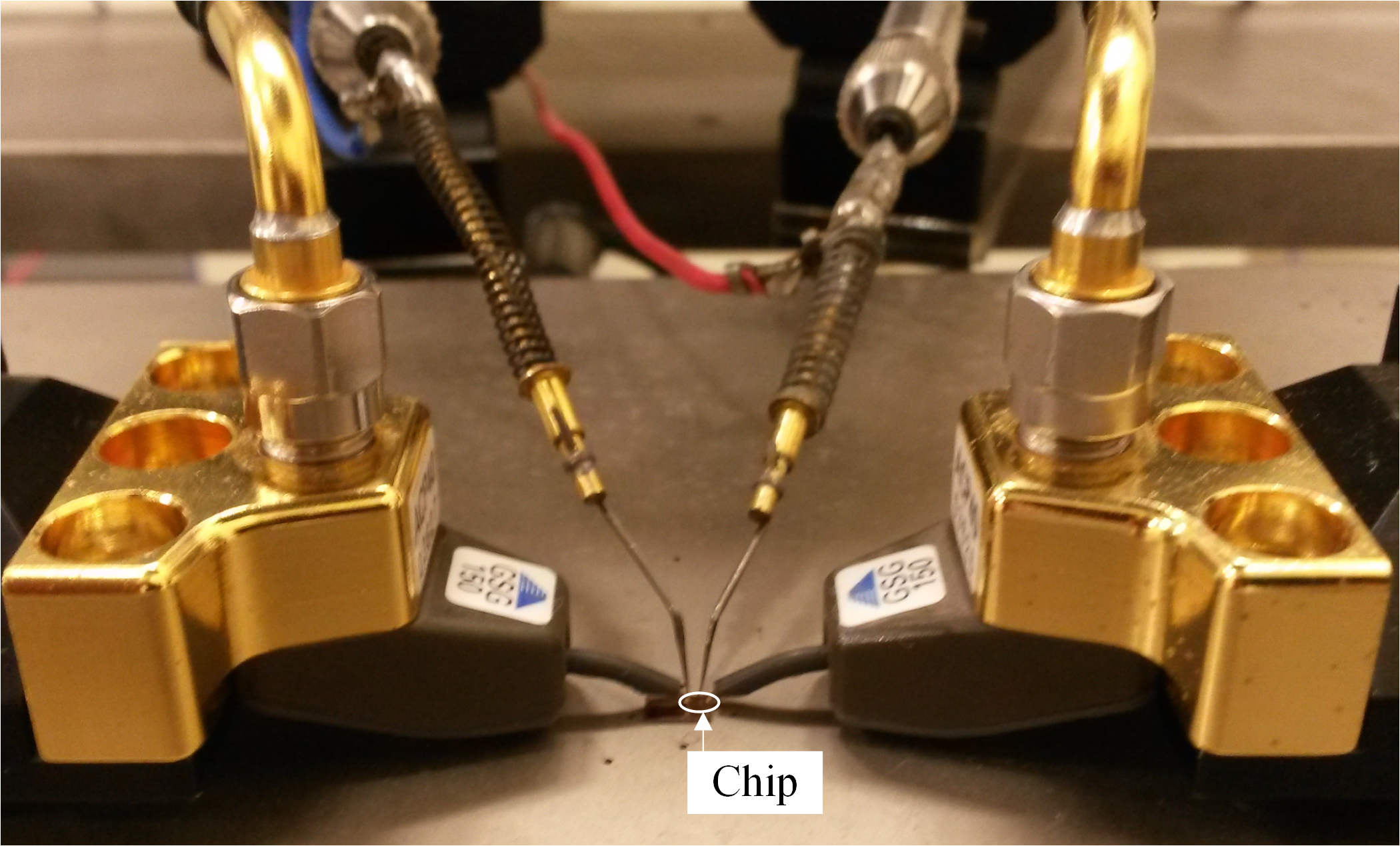}
 \caption{LNA measurement setup.}
 \label{fig6}
\end{figure}

Measurements were performed directly on-chip using a Cascade\textsuperscript{\textregistered} microprobe station (Figure~\ref{fig6}). S-parameters were measured with an HP\textsuperscript{\textregistered} 8510C network analyzer. To forward-bias the ESD protection diodes, a DC component of approximately 0.6~V was superimposed at the RF terminals by the analyzer itself. The minimum available RF excitation level in this analyzer is $-13$~dBm \cite{ref21}, which exceeds the maximum input level for which the LNA was designed and lies outside its linear range. This causes differences between the measurements and the S-parameter behavior under nominal IEEE 802.15.4 operating conditions.

\begin{figure}[!t]
 \centering
 \includegraphics[width=1\linewidth]{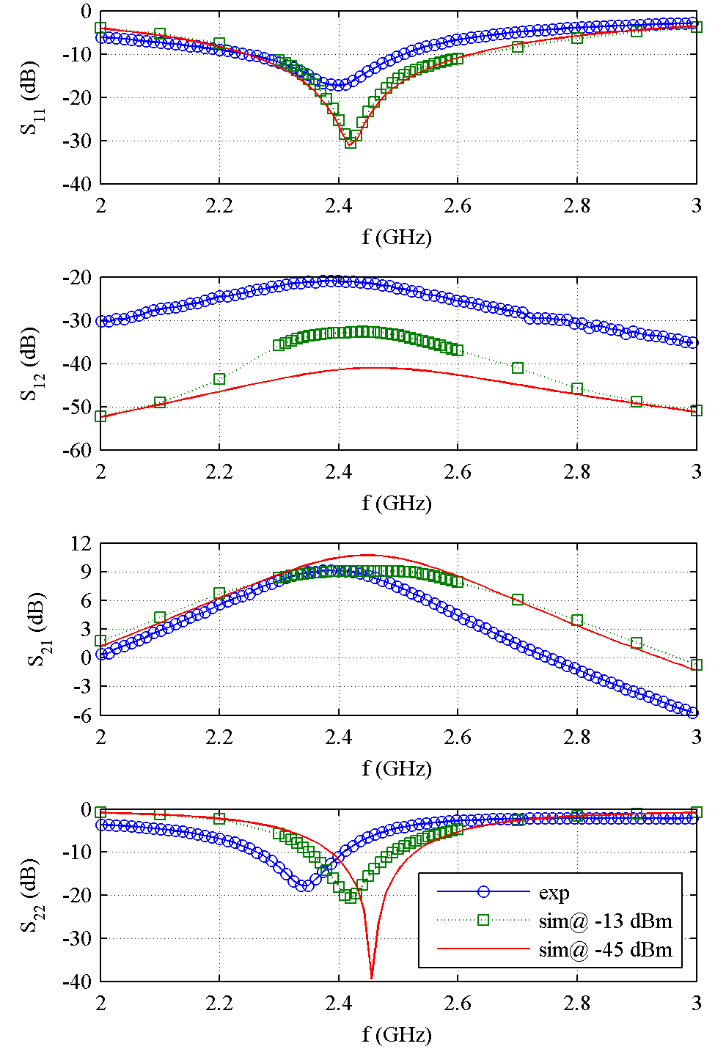}
 \caption{Frequency response of S-parameters: experimental vs. post-layout simulations.}
 \label{fig7}
\end{figure}

Figure~\ref{fig7} shows the frequency response of the S-parameters, comparing experimental results with post-layout simulations. For a more realistic comparison, post-layout simulations were performed at the same excitation used during measurement ($-13$~dBm). Since small-signal S-parameter simulations linearize the circuit around the operating point, they do not capture effects arising when the LNA is driven outside its linear range. Therefore, a large-signal simulation was employed to obtain S-parameters under nonlinear conditions (LSSP, Large-Signal Scattering Parameters \cite{ref22}).

Large-signal simulations confirm how the $-13$~dBm excitation alters the measured results, except for input matching ($S_{11}$). It is therefore reasonable to expect improved values for the other S-parameters when the LNA is driven with appropriate input levels: increased gain ($S_{21}$), improved reverse isolation ($S_{12}$), and reduced frequency shift in output matching ($S_{22}$). Although the measured input matching degrades with respect to simulation, it remains better than-10~dB across the band, thus meeting the specification. The gain response exhibits a steeper roll-off above the center frequency, suggesting parasitic capacitances higher than modeled; however, the gain at 2.45~GHz is only 0.5~dB lower than the simulated value at $-13$~dBm. Assuming a similar difference when the circuit operates with input levels below $-20$~dBm, the gain is expected to remain above 10~dB under normal operating conditions (the simulated gain at $-45$~dBm is 10.7~dB).

For the output matching, a shift toward lower frequencies is observed, which may be associated with process variations \cite{ref23}. The appearance of similar behavior in other published works—either at the same frequency with a different technology \cite{ref20} or at higher frequencies with a similar technology \cite{ref19}—suggests a non-random common cause. Thus, this deviation may also be due to parasitic capacitances larger than modeled (consistent with the gain behavior) and/or magnetic coupling among inductors \cite{ref19}. Reverse isolation likewise degrades with respect to simulations; this has been shown to occur primarily due to magnetic coupling between the drain, source, and gate inductors \cite{ref18,ref19}. Nevertheless, the measured value remains acceptable, and the LNA satisfies the usual stability criterion (e.g., $K>1$ and $\Delta<1$ \cite{ref24}).

\section{Conclusions}
\label{sec:conclusions}
This work presented the analysis, design, and measurement of a CMOS common-source LNA with inductive degeneration, contributing to RF IC research in Cuba. Designed for IEEE 802.15.4 at 2.4 GHz in 130 nm CMOS, post-layout simulations show 10.7 dB gain, 2.7 dB NF, 0.9 dBm IIP3, $S_{11}$ and $S_{22}$ better than -20 dB, and 505 $\mu$W consumption. Measurements confirmed simulations within expected deviations. The results and discussions presented can be used in future implementations to improve experimental performance in the desired frequency band.

\bibliographystyle{IEEEtran}
\bibliography{references}
\end{document}